\documentclass[12pt]{iopart}

\usepackage[OT2,T1]{fontenc}
\usepackage[english]{babel}
\usepackage[latin1]{inputenc}
\usepackage[charter]{mathdesign}
\usepackage[scaled=0.9]{helvet}
\usepackage{url,color}
\usepackage{enumerate}
\usepackage{verbatim}  

\usepackage{longtable}
\usepackage{tabularx}
\usepackage{multicol} 
\usepackage{multirow}
\usepackage{booktabs} 
\usepackage{ltxtable}
\usepackage{hhline}
\usepackage{bm}

\usepackage[final]{graphicx}
\usepackage{epsf}
\usepackage{wrapfig}
\usepackage{subfigure}
\usepackage{float}

\usepackage{caption}
\usepackage{mathrsfs}  

\usepackage[numbers,sort&compress]{natbib}

\usepackage{icomma}                        
\usepackage[pdfpagelabels,plainpages=false]{hyperref}  

\definecolor{uni}{rgb}{0.490,0.604,0.667}
\definecolor{ippblue}{cmyk}{1,0.45,0.04,0.}

\hypersetup{pdftitle={Double-resonant fast particle-wave interaction},
            pdfauthor={Mirjam Schneller},
            pdfsubject={Double-resonant fast particle-wave interaction},
            pdfcreator={\LaTeX2e},
colorlinks, linkcolor=black, urlcolor=blue, citecolor=green}

\usepackage{fancybox}

\begin{document}

\newcommand{\qp}{$q$ profile}
\newcommand{\bef}{$\beta_{\mathrm{f}}$}

\title[Double-resonant fast particle-wave interaction]{Double-resonant fast particle-wave interaction}

\author{M. Schneller$^1$, Ph. Lauber$^1$, M. Br{\"u}dgam$^1$, S. D. Pinches$^2$ and S. G{\"u}nter$^1$}
\address{$^1$ Max-Planck-Institut f{\"u}r Plasmaphysik, EURATOM Association, Boltzmannstr. 2, Garching
D-85748, Germany}
\address{$^2$ Culham Centre for Fusion Energy, Culham Science Centre, Abingdon, OX14 3DB, Oxfordshire, U.K.}
\ead{\mailto{mirjam.schneller@ipp.mpg.de}}

\begin{abstract}
In future fusion devices fast particles must be well confined in order to transfer their energy to 
the background plasma. 
Magnetohydrodynamic  instabilities like Toroidal Alfv{\'e}n Eigenmodes or core-localized modes such as 
Beta Induced Alfv{\'e}n Eigenmodes and Reversed Shear Alfv{\'e}n Eigenmodes, 
both driven by fast particles, can lead to significant losses. This is observed in many ASDEX 
Upgrade discharges.
The present study applies the drift-kinetic \textsc{Hagis} code with the aim of understanding
the underlying resonance mechanisms, especially in the presence of multiple modes with different
frequencies. Of particular interest is the resonant interaction of particles simultaneously with two
different modes, referred to as ``double-resonance''.
Various mode overlapping scenarios with different \qp s are considered.
It is found  that, depending on the radial mode distance, double-resonance is able to enhance growth 
rates as well as mode amplitudes significantly. Surprisingly, no radial mode overlap is necessary
for this effect. Quite the contrary is found: small radial mode distances can lead to strong nonlinear mode 
stabilization of a linearly dominant mode.
\end{abstract}

\section{Introduction}
    Fusion devices contain fast particle populations due to external plasma
    heating and (eventually) fusion-borne $\alpha$-particles.
    Fast particle populations can interact with global electromagnetic waves, 
    leading to the growth of MHD-like and kinetic instabilities -- e.g., Toroidicity
    Induced Eigenmodes (TAE) \cite{Cheng85, Cheng86}, Reversed Shear 
    Alfv{\'e}n Eigenmodes (RSAE) \cite{Berk01}
    or Beta Induced Alfv{\'e}n Eigenmodes (BAE) \cite{Heidbrink93, Turnbull93}.\\\\
    In this work, drift-kinetic fast particle simulations performed
    with the \textsc{Hagis} code (\cite{Pinches98,sip_phd}, shortly introduced in \sref{hagis}) are 
    carried out to obtain a deeper understanding of
    the processes in phase space that occur due to fast particle interaction
    with MHD modes. Of interest is the dynamics of wave-fast particle interaction in
    scenarios with two modes of different frequencies: what are the mode coupling mechanisms, and 
    what is the dependence on radial mode distance?
    \Sref{theory} gives an overview of the theoretical basis for
    these coupling mechanisms.
    \Sref{hagis} explains in brief, which ASDEX Upgrade reference scenario is selected and 
    how the MHD equilibrium is processed to give the basis for the \textsc{Hagis} simulations.
    In \sref{numstudy} numerical studies are described, investigating the stochastic threshold and the influence of $q$
    on the mode drive. It is found that the presence of a second mode can enhance growth rates as 
    well as mode amplitudes significantly. As this effect is based on particles that have resonance regions in phase
    space with both modes, it is referred to as ``double-resonance''. The double-resonant effect was expected
    to decrease with the radial mode distance, but it turned out, that the picture is more complicated.

\section{Theoretical Picture of Double-Resonance}\label{theory}
    Theory (e.g.\ Ref.\ \cite{Berk92}) predicts that conversion of free energy
    to wave energy is enhanced in a multiple-mode scenario, i.e., the interaction of multiple modes produces
    energy conversion rates higher than that which would be achieved with each mode acting independently. 
    This can be partially explained by the principle of gradient (of the radial particle distribution) driven
    mode growth -- according to\footnote{Throughout this work, $s$ refers to the radial coordinate as 
    the square root of the normalized poloidal flux: 
    $s=(\psi_{\mathrm{pol}}/\psi_{\mathrm{pol, edge}})^{1/2}$.}
    $\gamma \propto \nabla f(s)$ \cite{Fu89} -- which can be extended to 
    multiple modes \cite{Berk90-I, Berk90-II, Berk90-III,Berk92}.   
    This picture of \emph{\textbf{gradient driven double-resonance}} is based on the
    precondition that modes share resonances in the same phase space area.
    Through the resulting redistribution by each mode, a steeper gradient
    is produced at the other mode's position, enhancing its drive. The 
    overlapping of modes leads then to a much larger conversion of free energy to wave
    energy.\\
    However, this mechanism can only work if there is also \emph{spatial} mode overlap in 
    the radial direction. In Ref.\ \cite{mwb_phd} simulations were carried
    out, finding a double-resonant effect also without this precondition.
    Furthermore, a superimposed oscillation on the modes' amplitudes was observed,
    clearly indicating mode-mode interaction. The modes without radial overlap
    are then coupled radially through the particles' trajectories: A population
    of particles that shares resonances in phase space with both modes and passes 
    both modes' location at once, can transfer energy from one mode to the other \cite{mwb_phd}.
    Since the particle orbits are characterized by a certain width, it is not necessary that
    both modes have a radial overlap.
    In the following, this mechanism is called \emph{\textbf{inter-mode energy transfer}}:
    By damping one mode, particles gain energy $E$ and also toroidal momentum $P_{\zeta}$ 
    due to \cite{sip_phd}
    \begin{equation}\label{EPzeta}
        (E - \frac{\omega}{n}P_{\zeta}) = \mathrm{const.}
    \end{equation}
    As Alfv{\'e}nic mode frequencies are very low (compared to the particles' cyclotron frequency),
    momentum transfer dominates over  energy transfer.
    Since $P_{\zeta} \propto -\Psi$, particles gaining energy from the wave (i.e.\ damping it),
    are redistributed radially inwards, whereas particles losing energy to the
    wave (i.e.\ driving it), are redistributed radially outwards. The latter is the
    dominant process and is caused by the negative
    slope of the particles' radial distribution function.
    When passing through the second mode, the particles lose energy and toroidal momentum by 
    driving the mode.
    As there is no radial net drift in this mechanism, it can continue as long as the dominant
    mode is strong, making this the dominant process over other possible 
    combinations of mode-mode energy exchange. Particles that gain energy from
    both modes or lose energy to both modes soon leave the resonant phase space
    area.
    It is the exchange of energy between the modes that leads to the observed
    oscillation of their amplitudes: The mode receives energy in the rhythm of the
    particles' bounce frequency $\omega_{\mathrm{b}}$,
    which equals the beat frequency of both modes $\Delta \omega$ as shown in the following:
    Trapped particles with a bounce frequency $\omega_{\mathrm{b}}$ and toroidal 
    precession frequency $\omega_{\mathrm{tp}}$
    interact with MHD modes of a certain frequency $\omega$,
    if the \emph{resonance condition} \cite{porcelli94}
    \begin{equation}\label{rescond}
        \omega - n\omega_{\mathrm{tp}}-p\omega_{\mathrm{b}} \approx 0
    \end{equation}
    is fulfilled, where $n$ is the toroidal mode 
    harmonic and $p$ the particles' bounce harmonic. 
    For double-resonance, this resonance condition has to be fulfilled for both modes '1' and '2' 
    simultaneously, leading to 
    \begin{equation}\label{doublerescond}
      \omega_{\mathrm{tp}}(n_{1}-n_{2})+\omega_{\mathrm{b}}(p_{1} - p_{2}) 
      = \omega_{\mathrm{1}}-\omega_{\mathrm{2}} \equiv \Delta \omega
    \end{equation}
    For the simplest case, the toroidal mode numbers $n$ are considered equal, 
    and $p_{1} = 1$, $p_{2} = 0$. 
    This leads to $\omega_{\mathrm{b}} = \Delta \omega$.
    Indeed, the lowest bounce harmonics $p=0, \pm 1$ are the most relevant ones; 
    however, the issue of different toroidal mode numbers $n$ will be discussed later.

\section{Using the Hagis Code with an ASDEX Upgrade Plasma Equilibrium}\label{hagis}
    The numerical investigations presented in this work are performed with 
    the \textsc{Hagis} Code \cite{Pinches98, sip_phd},
    a nonlinear, drift-kinetic, perturbative Particle-in-Cell code (current release 12.05).
    \textsc{Hagis} models the interaction between a distribution of energetic particles and a set
    of Alfv{\'e}n Eigenmodes. It calculates the linear growth rates as well as the nonlinear behavior 
    of the mode amplitudes and the fast ion distribution function that are determined by 
    kinetic wave-particle nonlinearities. It is fully
    updated to work with MHD equilibria given by the recent \textsc{Helena} \cite{Huysmans91} version.\\
    \begin{minipage}{1\textwidth}
      \begin{minipage}{0.45\textwidth}
        \begin{figure}[H]
           \hspace{1cm}\includegraphics[width=0.7\textwidth]{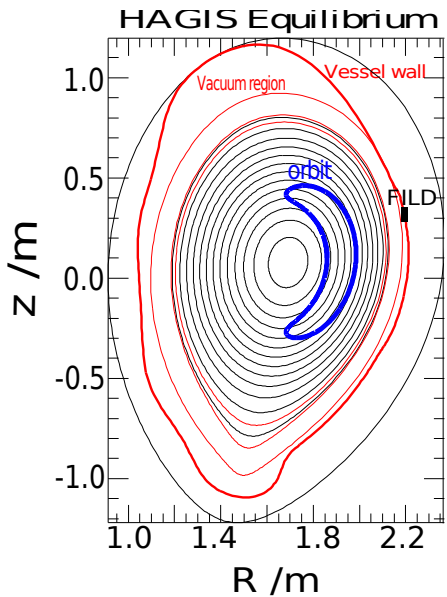}
           \caption{\itshape Hagis equilibrium with vacuum region (red), 
             representative banana orbit (blue) and FILD position (black square).}
           \label{hagis_equilibrium}
        \end{figure}
        \vfill
      \end{minipage}
      \hfill
      \begin{minipage}{0.45\textwidth}
        \begin{figure}[H]
          \includegraphics[width=0.9\textwidth, height=4.7cm]{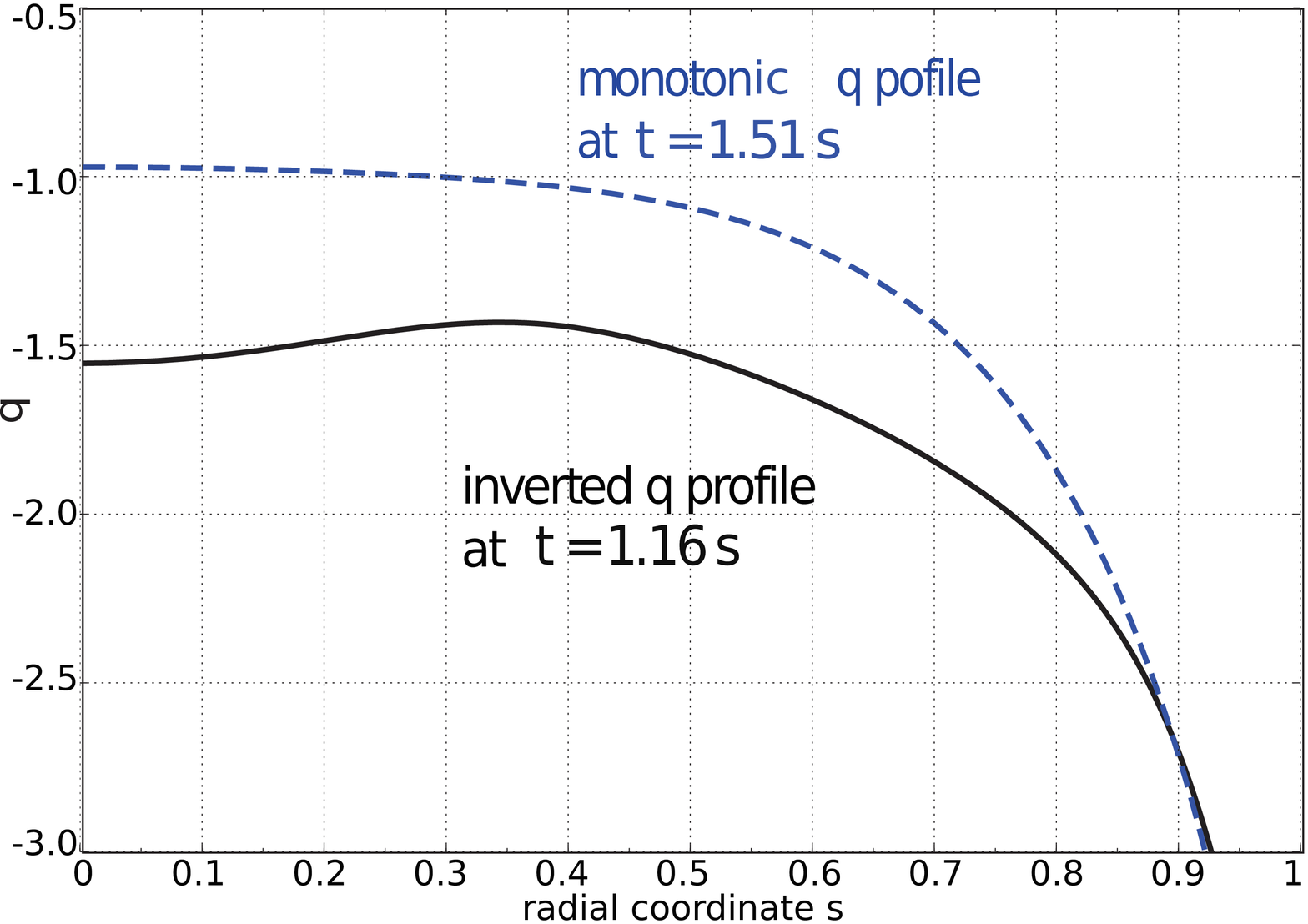}
          \caption{\itshape \qp\ according to experimental measurements in
            AUG discharge \#23824 at different time 
          points: $t=1.16$\ s (black solid line)  and  $t=1.51$\ s (blue dashed line):
          the \qp s differ in the shape (the earlier one is inverted, the later one monotonic), but
          also in the absolute values (the inverted \qp\ has higher absolute $q$ values).}
          \label{q23824}
        \end{figure}
      \end{minipage}
    \end{minipage}
    The plasma equilibrium (\fref{hagis_equilibrium}) for \textsc{Hagis} is based 
    on the \textsc{Cliste} code \cite{Carthy99, Carthy12},
    then transformed via \textsc{Helena} to straight field line coordinates and
    to Boozer coordinates by \textsc{Hagis}.\\
    The data for the MHD equilibria of all simulations presented in the following originate from the
    ICRH minority heated ASDEX Upgrade (AUG) discharge \#23824, at time  $t=1.16$\ s 
    or $t=1.51$\ s.
    At the earlier time point, the \qp\ is slightly inverted (\fref{q23824}, black solid line),
    whereas at the later time point, it is monotonic, with lower absolute values (\fref{q23824}, 
    blue dashed line).\\
    The plasma equilibrium, and in particular the \qp\ is determined by Alfv{\'e}n spectroscopy of
    the RSAEs:
    magnetic pick-up coil data and soft X-ray emission measurements are used to determine the
    \qp's minimum value and location.\\
    This particular discharge was chosen due to the availability of especially detailed 
    experimental data concerning fast particle-mode interaction. However, a detailed comparison between 
    numerical and experimental results goes beyond the scope of this work, but is the
    object of current and future
    investigation. In this work, a numerical study is presented, that is based on an AUG reference case:
    the \qp\ as well as the mode frequencies, harmonics and widths are chosen according to
    experimental observation. However, significant abstractions have also been made: 
    the mode structures are chosen analytically, as well
    as the particle distribution functions. Furthermore, mode saturation is based completely on
    the depletion of the particle distribution function, no collisions, nor turbulence are taken
    into account, and there is no source term for energetic particles.

\section{Numerical Study on Double Mode Resonance}\label{numstudy}
  In this section, a numerical study is presented that examines
  under which circumstances mode-particle
  interaction can take place. A special focus is set on the interaction 
  with multiple modes of different frequencies.

  \subsection{Simulation Conditions}
\begin{figure}[H]
        \centering
        \includegraphics[width=0.4\textwidth, height=3.5cm]{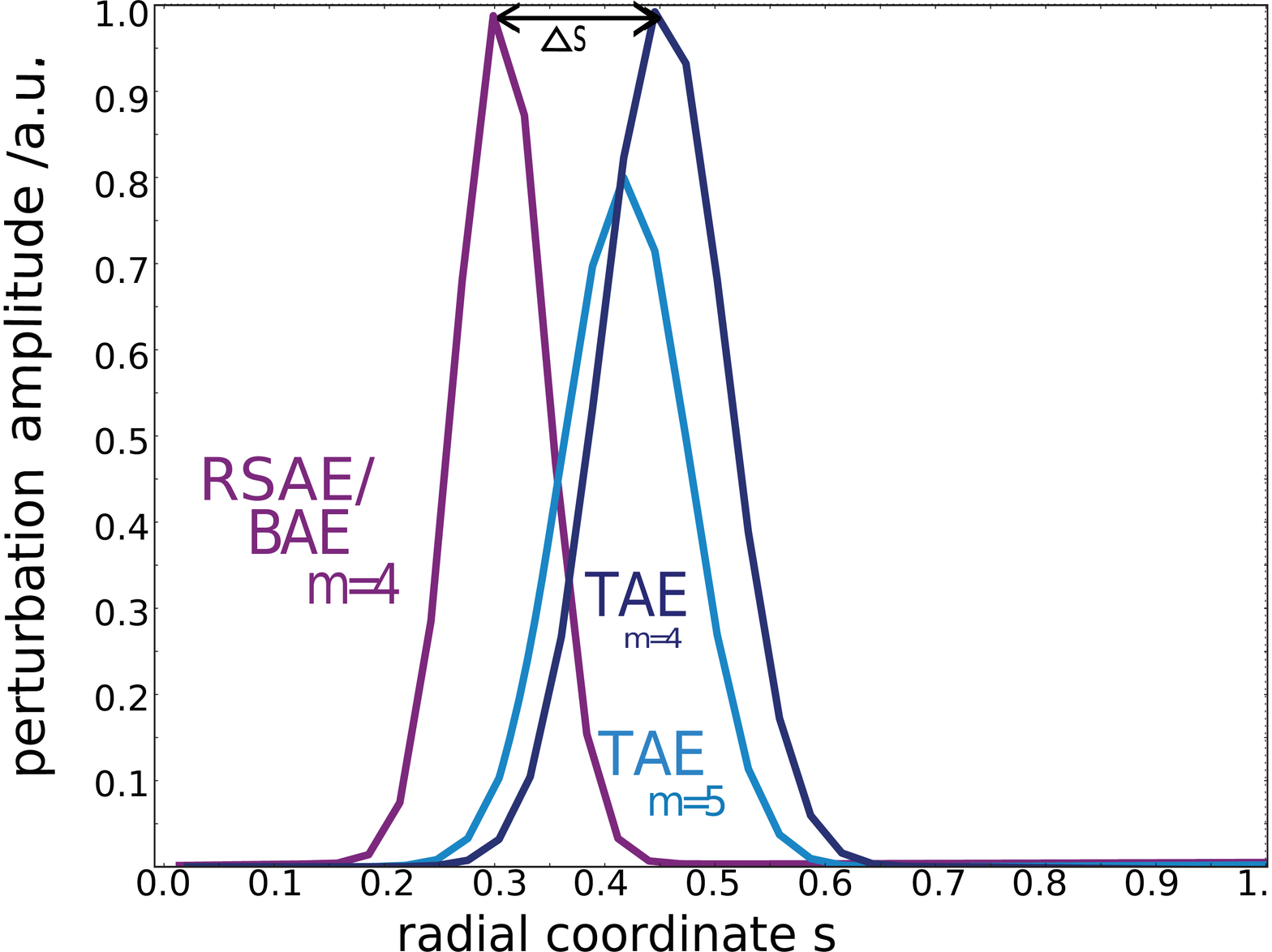}
        \caption{\itshape Analytical, gauss-shaped perturbation as used in the simulations.}
        \label{pert_exam}
     \end{figure}

    As the question of understanding double mode resonance is very fundamental, 
    the simulations were performed under quite simple, but still realistic physical conditions:
    the volume averaged fast particle beta is chosen to be \bef\ $= 1\%$. 
    To avoid different mode drive at different radial mode positions
    only due to a steeper gradient in the distribution function, a radial particle distribution $f(\psi)$ 
    with constant gradient is chosen. 
    As energy distribution function $f(E)$ a slowing down function \cite{Gaffey76} is used:
    \begin{equation}\label{slowdown}
      f(E) = \frac{1}{E^{3/2}+E^{3/2}_{\mathrm{c}}}\mathrm{erfc}\left(\frac{E-E_0}{\Delta E}\right)
    \end{equation}
    ($\Delta E=0.1499$\ MeV, $E_{\mathrm{c}}=0.1934$\ MeV, $E_0=1.0$\ MeV).
    The particles are distributed isotropically in pitch angle (as
    e.g., fusion-borne $\alpha$ particles would be). 
    As MHD perturbations, analytic, gauss shaped functions are used (see \fref{pert_exam}), 
    without background damping. The mode frequencies are chosen to
    match experimental data: at both time points within the considered discharge \#23824, 
    a high frequency TAE with $120$\ kHz is found, as well as
    a lower frequency mode at $55$\ kHz. Ref.\ \cite{Lauber09} describes 
    their mode as a RSAE at $t= 1.16$\ s
    and as BAE at $t= 1.51$\ s. The widths of the modes are based on the MHD eigenfunctions, calculated
    numerically with the linear eigenvalue solver \textsc{Ligka} \cite{Lauber07}.
    Convergence tests indicate that simulations with 120000 markers are
    sufficient, leading to computational costs of a few hundred CPUh for each simulation.

  \subsection{Simulation Results}
\begin{figure}[H]
      \centering
      \includegraphics[width=0.6\textwidth]{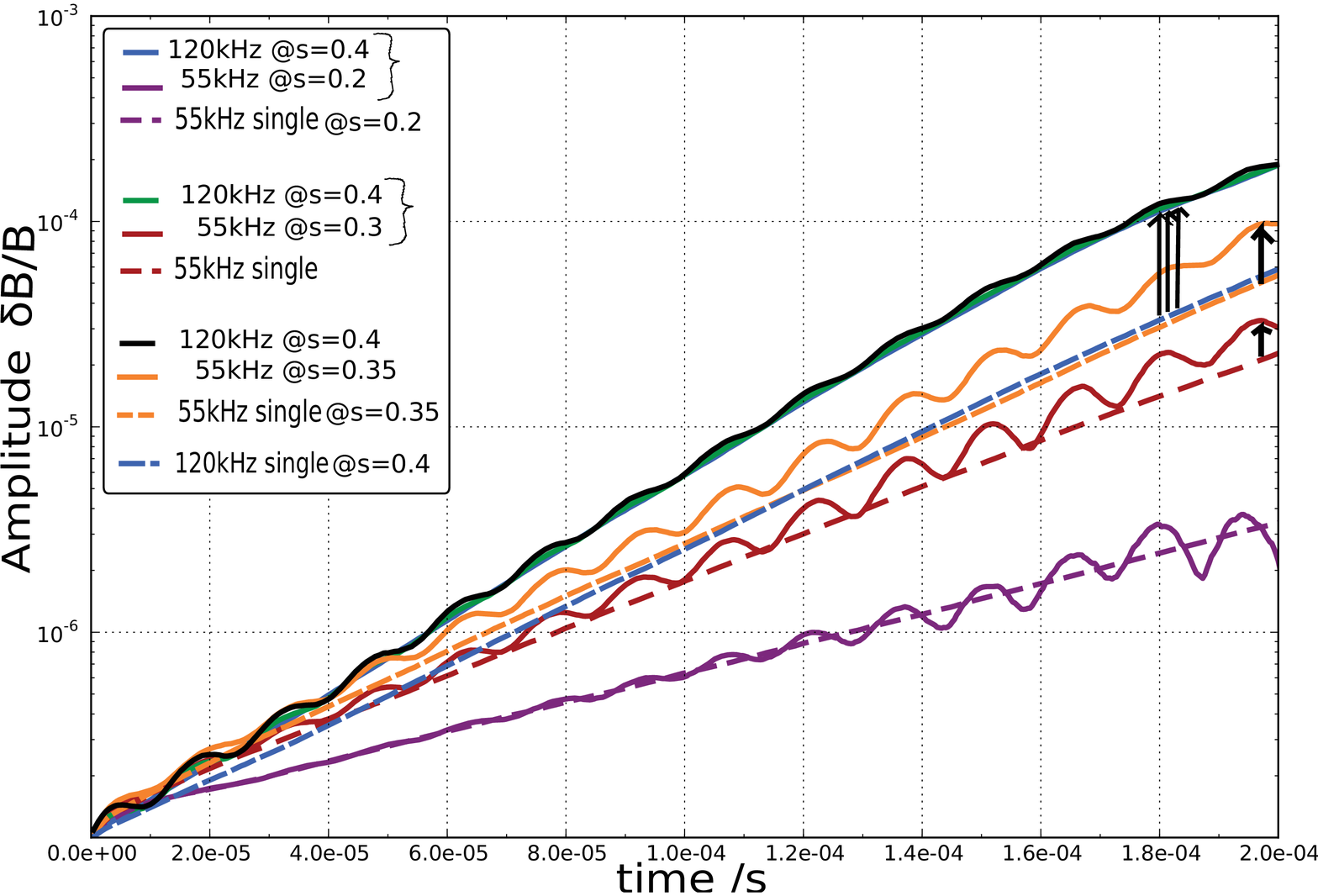}
      \caption{\itshape Mode amplitude evolution in linear growth phase for different 
        multi-mode (solid lines) scenarios (for the MHD equilibrium at $t=1.16$\ s),
        differing in radial mode distances $\Delta s$. For comparison, the single mode
        simulations are depicted as dashed lines. In all cases, the TAE 
        (120\ kHz at $s=0.4$, dark colors)
        grows stronger in the double mode scenario, the RSAE 
        (55\ kHz at various radial positions, reddish colors) in contrast, 
        benefits from the double mode case better, if $\Delta s$ is small enough.}
      \label{run0161+0162+0163+0165+0166+0175_linamp}
    \end{figure}

\begin{figure}[H]
      \centering
      \includegraphics[width=0.6\textwidth,height=5.5cm]{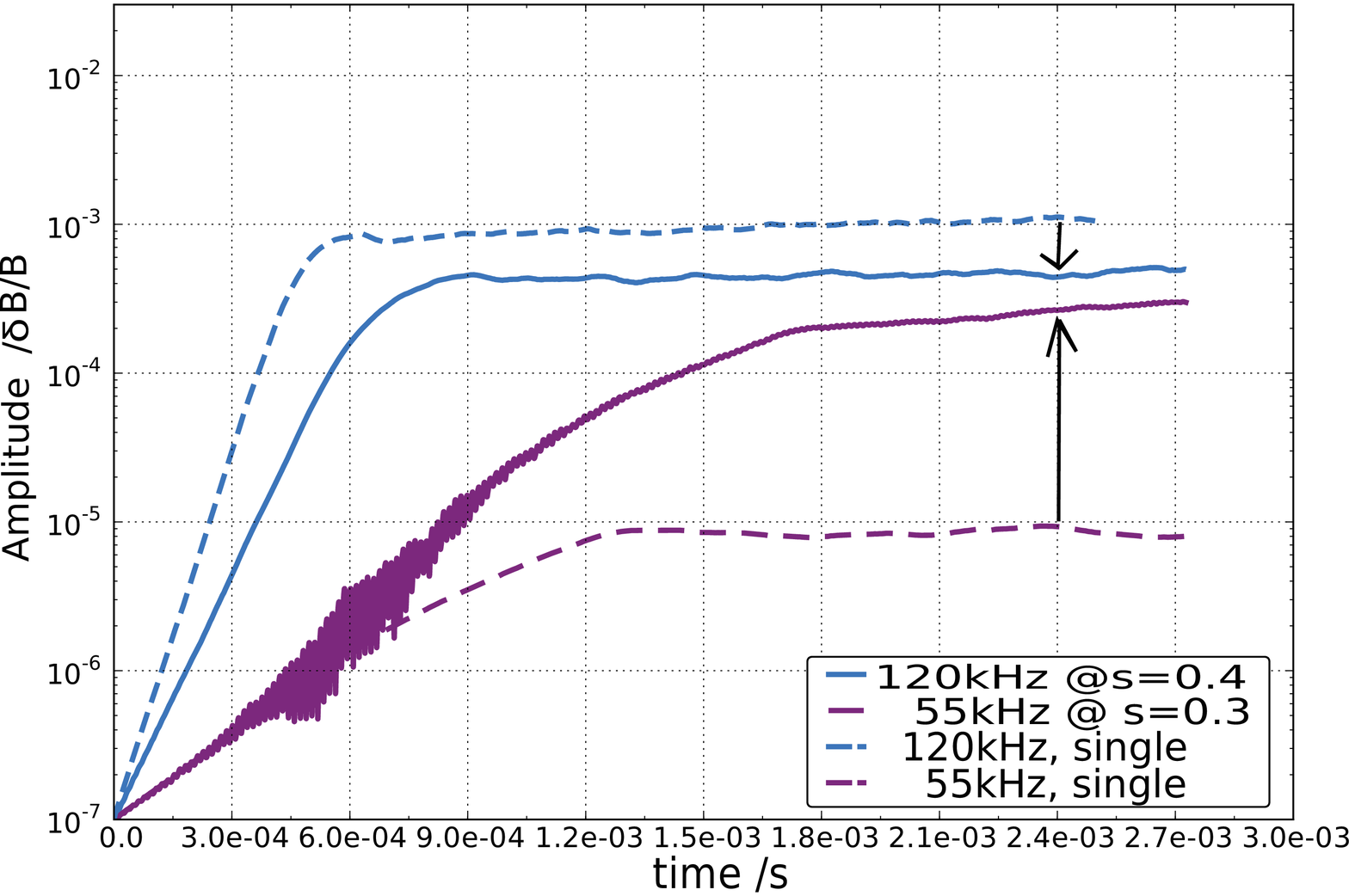}
      \caption{\itshape Mode amplitudes over time in double mode (solid lines) versus single 
        mode (dashed) simulation with low growth rates (for the MHD equilibrium 
        at $t=1.51$\ s,  \bef\ $= 2.5\%$ with 
        BAE (pink) at $s=0.3$ and TAE (blue) at $s=0.4$). Here, the 
        redistribution is too weak to lead to significant 
        gradient driven double-resonance. Inter-mode energy 
        transfer (from TAE to BAE) dominates.}
      \label{run0186+0187+0188_amplitude}
    \end{figure}

    Multi-mode simulations are carried out with different radial distances $\Delta s$ between
    the Alfv{\'e}nic modes.
    It is helpful to look at the different stages within the simulation individually:
    although all simulations are performed including all nonlinearities, the nonlinear
    effects are small at the beginning of each simulation, leading to an amplitude
    growing according to $\propto \mathrm{exp}(\gamma t)$ with $\gamma$ the so-called
    linear growth rate.
    \paragraph{The linear regime}
    The modes' amplitudes in this linear regime, as depicted in 
    \fref{run0161+0162+0163+0165+0166+0175_linamp}
    confirm already some of the results presented in \cite{mwb_phd}:
    at least one mode grows faster in the double mode scenario compared 
    to the single mode simulation (the TAEs in \fref{run0161+0162+0163+0165+0166+0175_linamp}
    dark colors).
    In most cases, both modes have larger growth rates in the double mode scenario than 
    in the single mode simulation. Furthermore, it is usually the radially outer mode that 
    is driven most strongly due to the double mode resonance.
    This is because \textit{gradient driven double-resonance} is most effective in 
    driving the outer mode: through resonant particle redistribution, one mode produces steeper
    gradients at the other mode's position. Through the inner mode, particles are 
    redistributed towards the outer position and supply it with more (resonant) particles,
    that can transfer their energy to the wave. At the inner mode's position, in contrast, 
    the particle population decreases, and with it, the energy transfer to the wave.\\
    However, in simulations with low growth rates 
    (see linear phase of \fref{run0186+0187+0188_amplitude}), 
    the stronger -- and that is mostly the outer mode -- was enhanced
    in mode drive less effectively than the subdominant one, or even weakened  
    compared to the single mode case. In these cases, the \textit{inter-mode energy exchange}
    is the prevailing mechanism, conducting energy from the stronger
    to the weaker mode.\\
    A superimposed oscillation is visible, strongest if the 
    modes' growth rates are relatively
    different from each other. However, if they differ too much, 
    the oscillation becomes jagged, due to
    the large impact of the dominant mode on the subdominant one.
    The oscillation frequency matches the beat frequency of both
    modes $\Delta \omega = \omega(\mathrm{TAE}) - \omega(\mathrm{RSAE}) = 65~$kHz, 
    a hint for working inter-mode energy transfer.\\
    As stated in Ref.\ \cite{mwb_phd}, both mode numbers $n$ have to be equal to lead
    to the oscillatory behavior, i.e.\ to double mode resonance, as
    will be discussed later.\\\\
    In the course of each simulation, nonlinear effects become more and more important,
    leading to a mode amplitude evolution different from $\propto \mathrm{exp}(\gamma t)$.
    The beginning of the nonlinear regime is observed, when the mode amplitude changes 
    from evolving linearly (in a semilogarithmic plot) to saturating at a certain level.\\

\begin{figure}[H]
      \centering
      \subfigure[\itshape during resonant phase]{\includegraphics[width=0.42\textwidth, height=3.0cm]{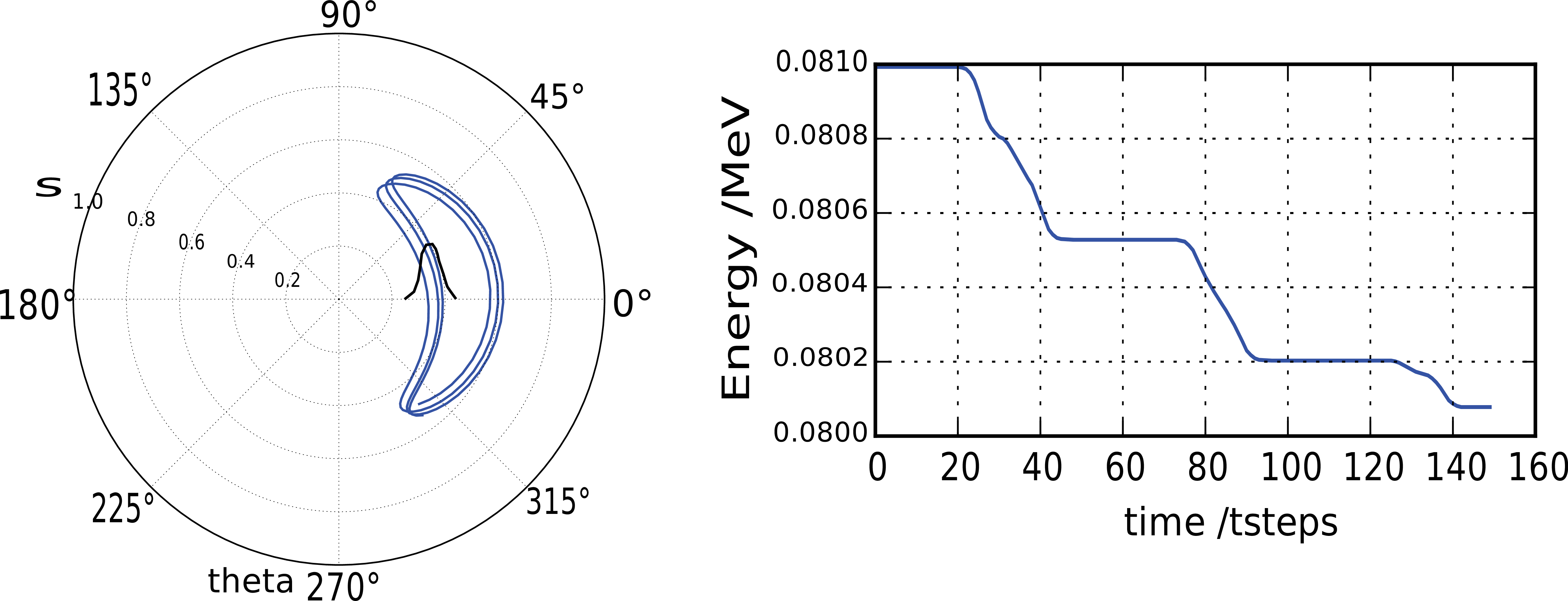}}
      \hspace{0.8cm}\subfigure[\itshape during stochastic phase]{\includegraphics[width=0.42\textwidth, height=3.0cm]{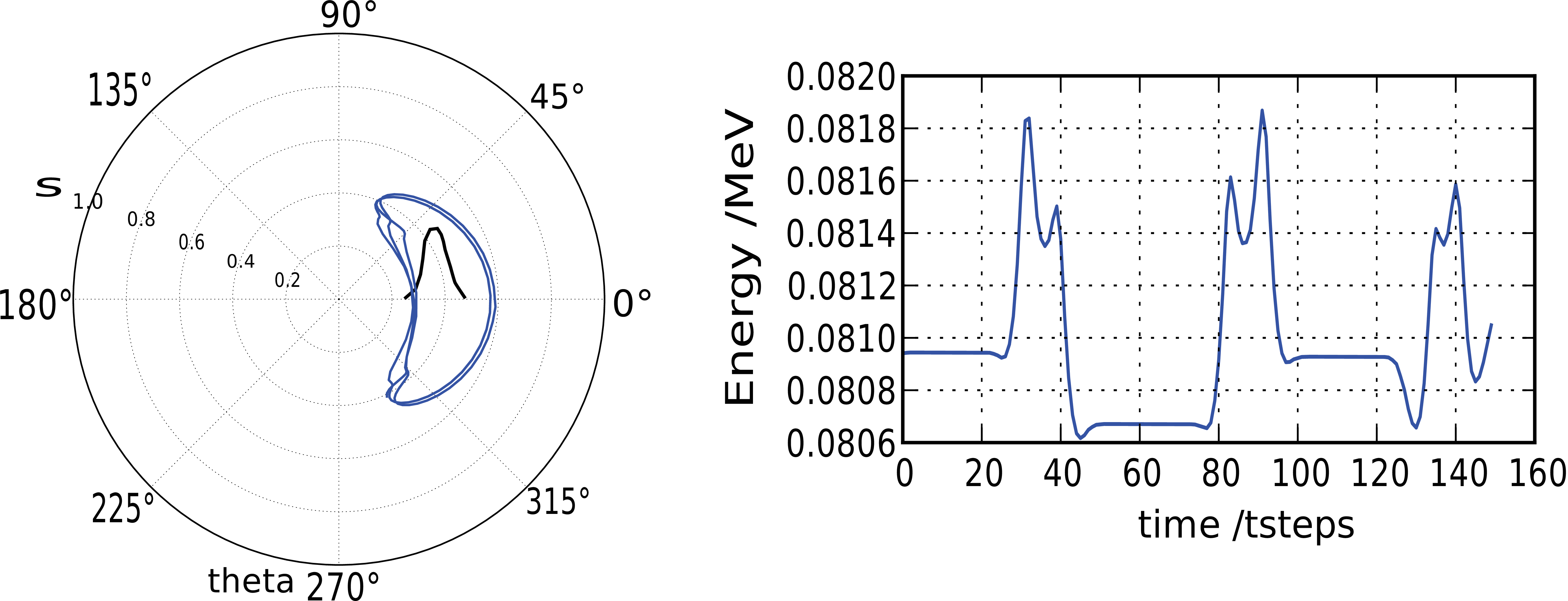}}
      \caption{\itshape Representative particle orbit (left, blue line) and energy evolution (right) 
        before (a) and after (b) the stochastic threshold is
        reached: During the early saturation phase, the orbit is banana-shaped and shifted outwards, with
        the particle losing its energy. In the stochastic regime, the orbit becomes distorted
        and the particle's energy loss occurs in bursts. The black line in the left plots indicates radial
        envelope of the MHD mode.}
      \label{run0175_orbit}
    \end{figure}

\begin{figure}[H]
      \centering
      \includegraphics[width=0.6\textwidth, height=5.5cm]{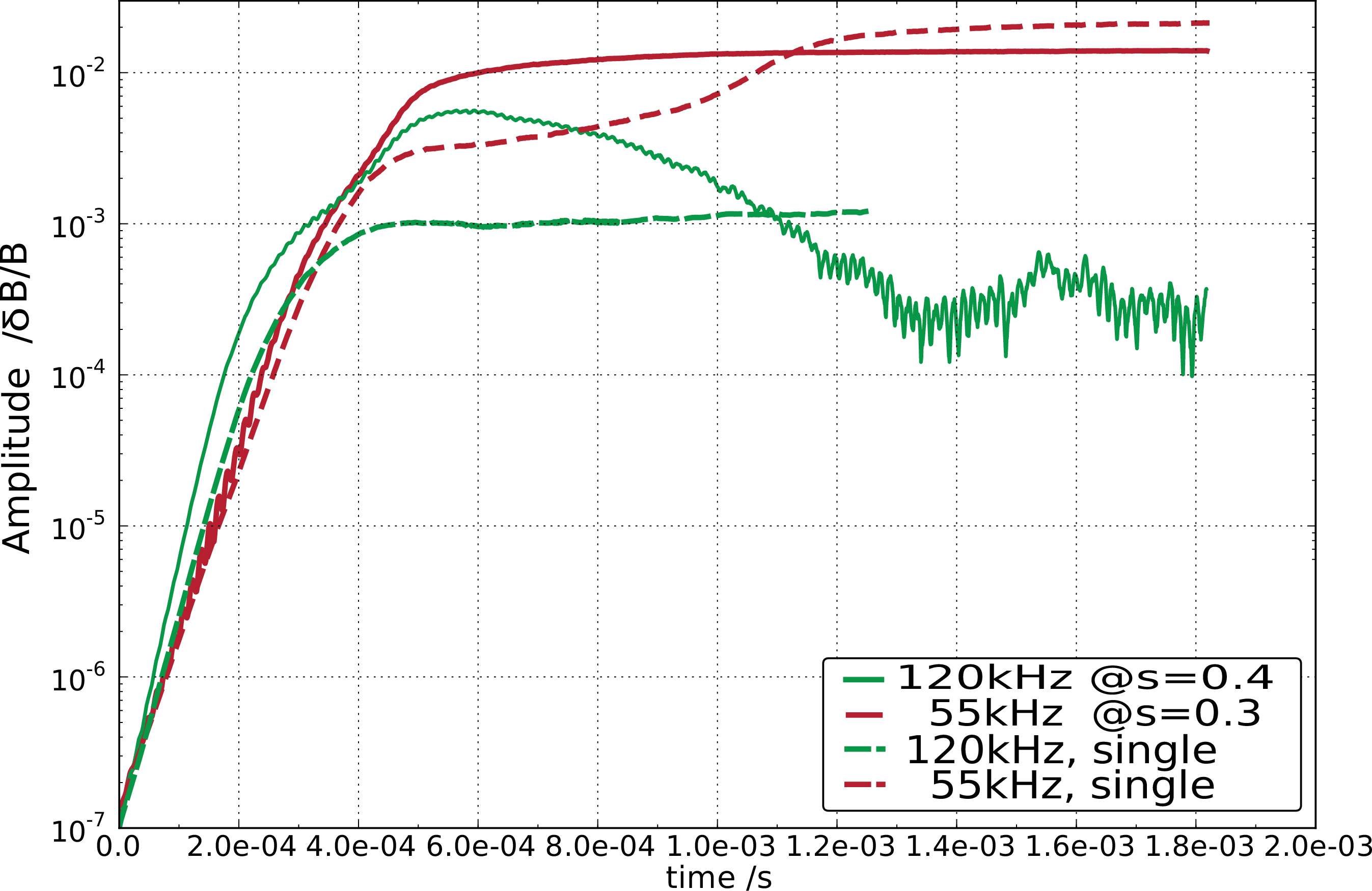}
      \caption{\itshape Mode amplitudes evolution in double mode (straight lines) 
        versus single mode (dashed) simulation, 
        with a RSAE (red) at $s=0.3$ and a TAE (green) at $s=0.4$ for 
        the MHD equilibrium at $t=1.16$\ s. Here, the TAE is destroyed
        in the double mode case.}
      \label{run0162+0165+0167_ampl}
    \end{figure}

\begin{figure}[H]
      \centering
      \includegraphics[width=0.4\textwidth]{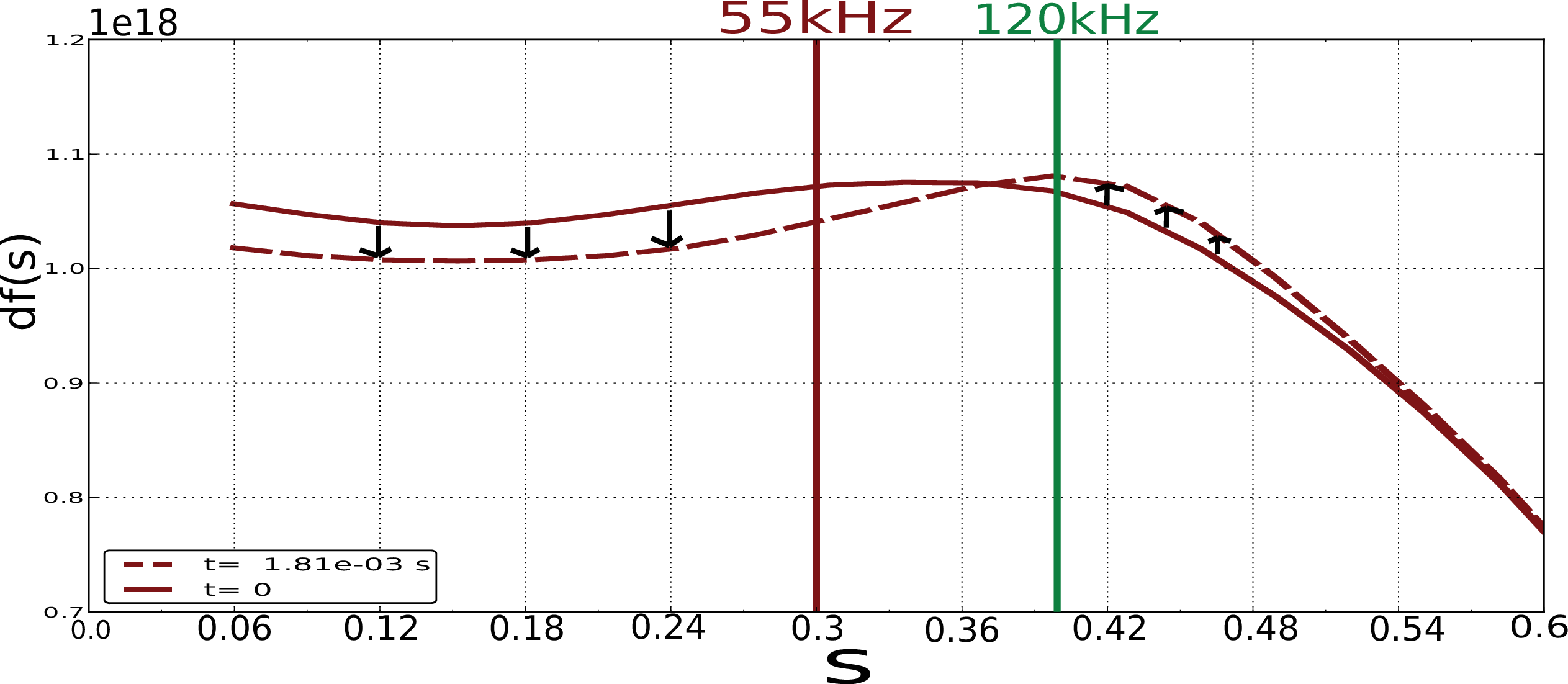}\\
      \includegraphics[width=0.4\textwidth]{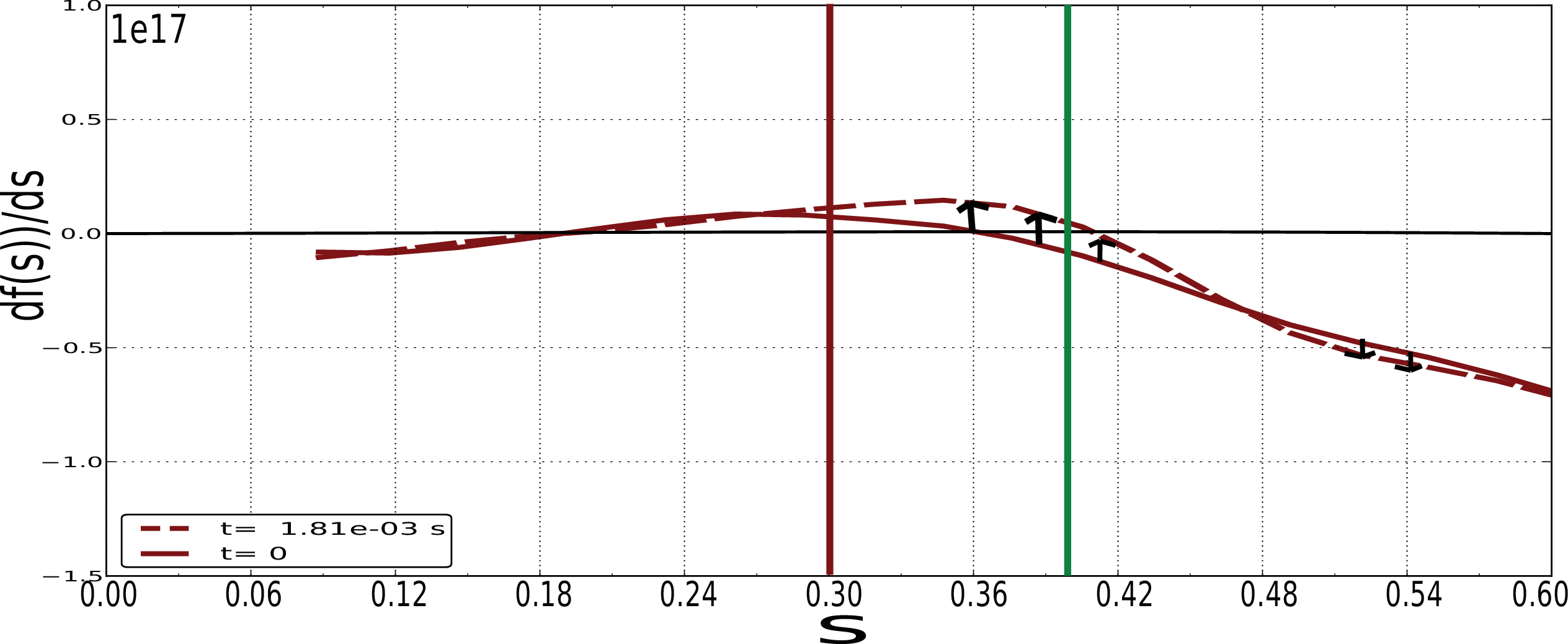}
      \caption{\itshape Temporal evolution (early: solid, late: dashed) of the distribution 
        function $f(s)$ in the single mode simulation 
       ($t=1.16$\ s equilibrium) with a RSAE at $s=0.3$ (upper). 
       Lower: the gradient $\mathrm{d}f(s)/\mathrm{d}s$ at $s=0.4$
       shows a transition from negative values (straight) to positive values (dashed), 
       leading to a very low amplitude of the
       TAE mode at that position in the double mode simulation.}
      \label{run0165_disb}
    \end{figure}

\begin{figure}[H]
      \centering
      \subfigure[\itshape Growth rates enhancement]{\includegraphics[width=0.37\textwidth,height=4cm]{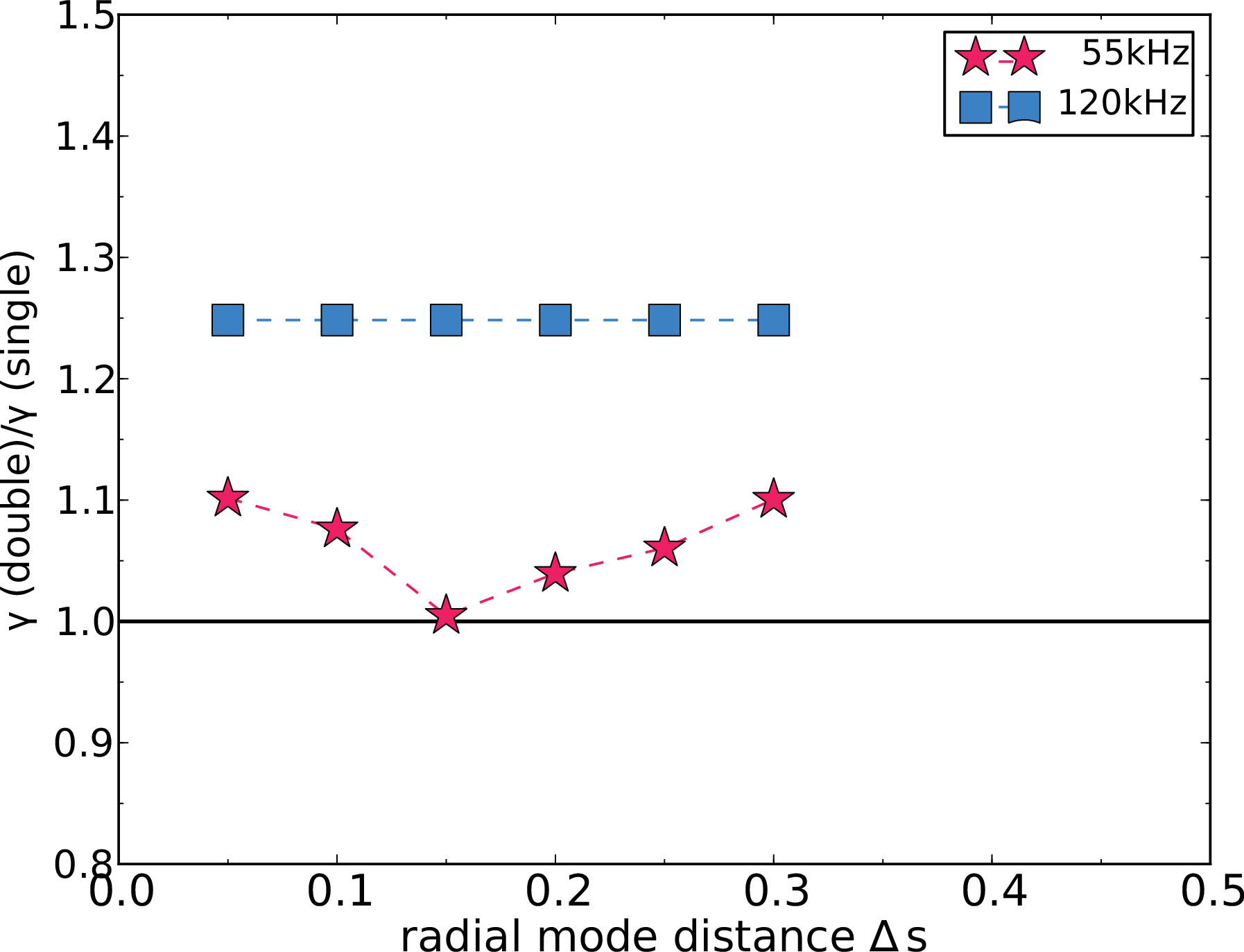}}
      \hspace{1.0cm}\subfigure[\itshape Amplitudes enhancement (at $\approx 300$ TAE periods)]{\includegraphics[width=0.37\textwidth,height=4cm]{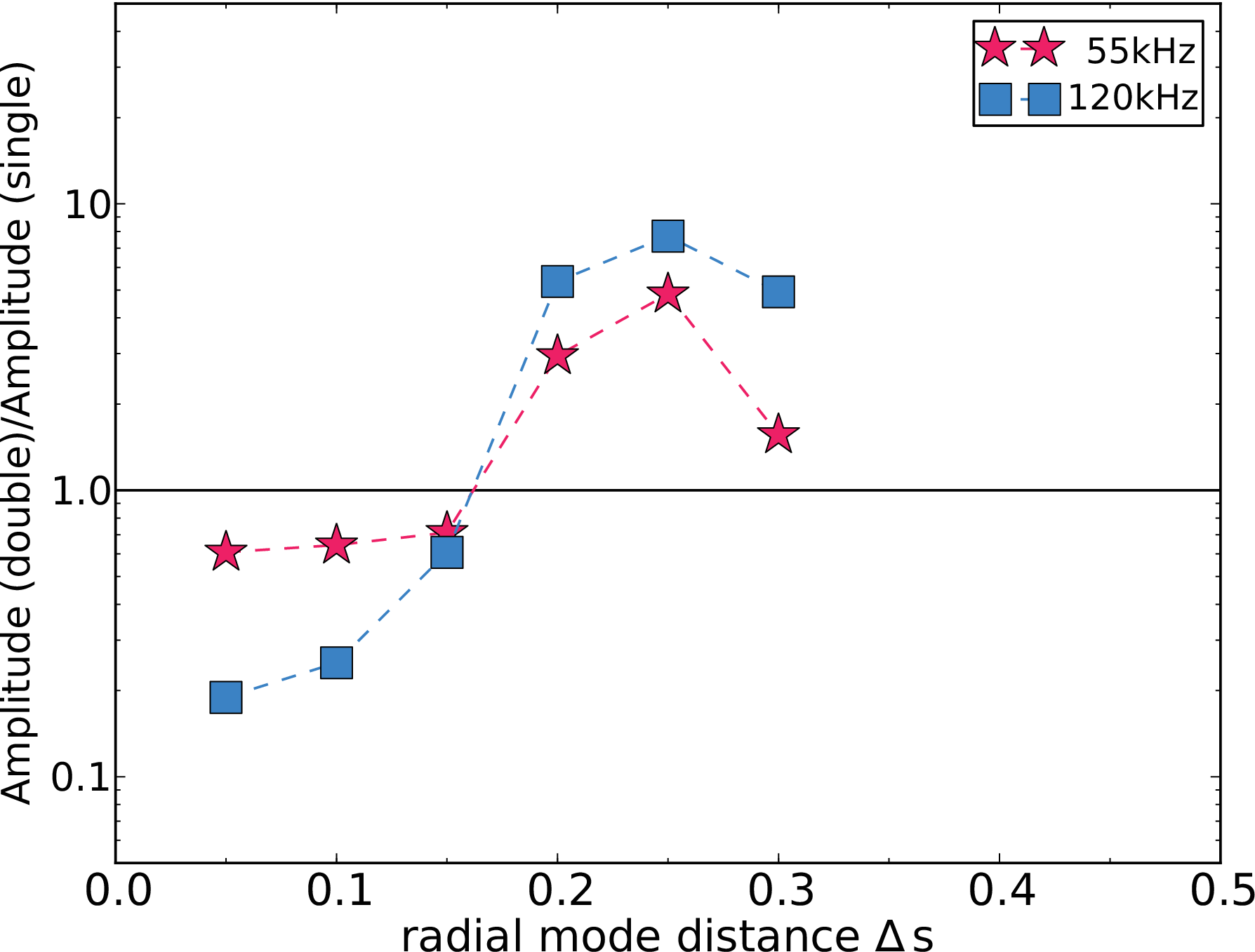}}
      \caption{\itshape Mode evolution scanned over the radial mode distance $\Delta s$ in the
        inverted \qp\ case. Depicted is the $\Delta s$ dependence of the ratios 
        of growth rates (a) and amplitudes (b) in double mode 
        simulations over those from single mode simulations. Red: RSAE, blue: TAE.
        One can clearly see that larger radial mode distances lead to higher amplitudes,
        whereas amplitudes are even lower than in the single mode case for $\Delta s \le 0.15$.
        The linear growth rates, however, are higher than in the single mode simulation throughout the
        $\Delta s$ range. The RSAE growth rate experiences a small drop at $\Delta s \approx 0.15$.}
      \label{deltas-scan}
    \end{figure}

    \paragraph{The nonlinear regime}
    The effect of the double-resonance in the nonlinear regime is clearly visible  
    in \fref{run0163+0166+0167_amplitude} ($\Delta s =0.2$): 
    \begin{figure}[H]
      \centering
      \includegraphics[width=0.6\textwidth, height=5.5cm]{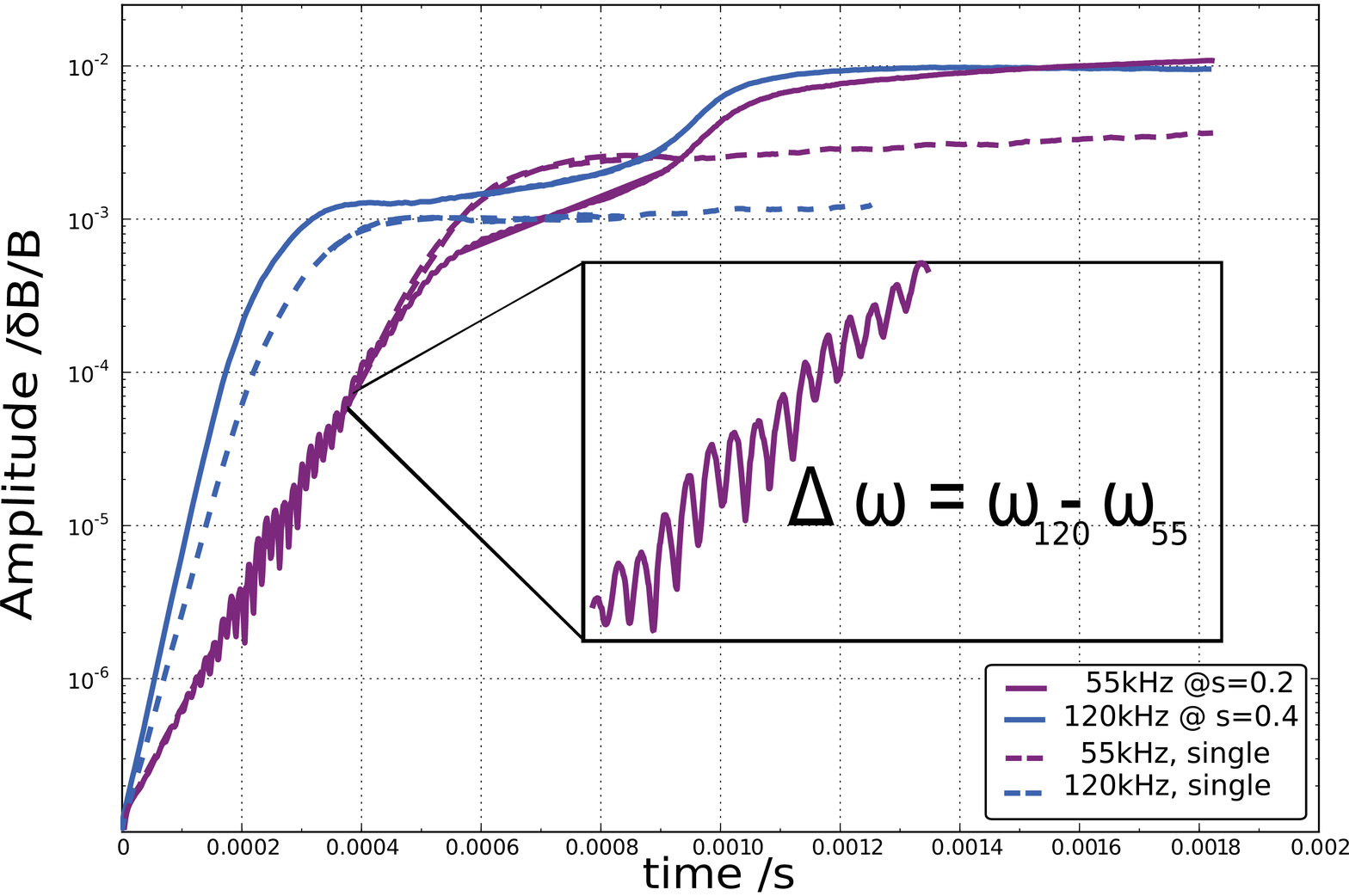}
      \caption{\itshape Mode amplitudes evolution in double mode (solid lines) 
        versus single mode (dashed) simulation, 
        with a RSAE (pink) at $s=0.2$ and a TAE (blue) at $s=0.4$ for the MHD 
        equilibrium at $t=1.16$\ s. In the double mode case, one
        can clearly see a superimposed oscillation and higher saturation levels.}
      \label{run0163+0166+0167_amplitude}
    \end{figure}
    in the double mode scenario, the modes not only grow faster (in the case of the TAE), 
    but also saturate at a higher level compared to the single mode case (by about a factor of five to ten). 
    This is due to the orbit \textit{stochasticization} at 
    $\delta B/B \approx 2 \cdot 10^{-3}$
    already discovered in single mode simulations: The stochastic
    threshold \cite{Berk92} is only reached in the double mode scenario, not in the single mode case. 
    To visualize the effect of stochasticization, an representative particle orbit and the particle's energy 
    evolution \textit{before} and \textit{after} the stochasticization sets in is shown 
    in \fref{run0175_orbit}.
    The picture is not as simple in the second example considered 
    ($\Delta s=0.1$, \fref{run0162+0165+0167_ampl}): 
    at the beginning of the nonlinear regime, 
    double-resonance is effective -- mode amplitudes are higher by a factor of five compared
    to the single mode simulation. At a later time point in the simulation 
    however, the situation changes
    completely: the TAE amplitude vanishes and the RSAE saturates at a lower level compared to
    the amplitude it reached in the single mode scenario during stochasticization. This scenario
    was found frequently for close radial mode distances, and shows that a linearly dominant
    mode can be stabilized nonlinearly in a multi-mode scenario.\\
    The reason for this behavior can be found in the radial particle redistribution.
    To find out the redistribution caused solely by the RSAE, the single mode
    simulation of that mode has to be considered: the RSAE leads to a radial particle redistribution 
    as depicted in \fref{run0165_disb}a). This redistribution leads 
    to a flattened gradient (\fref{run0165_disb}b) at $s=0.4$, 
    the position of the TAE in the 
    the double mode case of \fref{run0162+0165+0167_ampl}. The reason for the
    destruction of the TAE in this simulation is therefore its radial location too close to the RSAE.
    However, at all positions $s > 0.45$, the gradient becomes steeper in the
    course of the single mode simulation. Indeed, simulating the RSAE with a TAE in this 
    radial range leads to strong TAE drive.
    A scan over the radial mode distance reveals an effective double-resonance as shown in 
    \fref{deltas-scan}: depicted are the ratios of the linear growth rates (a) and the amplitudes 
    (b) in the double mode case vs.\ the single 
    mode case over the radial mode distance $\Delta s$. 
    The amplitude level was compared after $\approx 300$ TAE periods ($=~2.5$\ ms) of 
    simulated time. This time is sufficient for the single amplitudes to saturate,
    but still significantly below energy slowing down time. Therefore, the fact that
    no source term is implemented in the code does not disturb the physical picture.
    One can see that the growth rates of
    both the TAE and the RSAE are enhanced in all double mode cases compared to the single mode ones.
    However, the growth rate of the outer TAE is enhanced most strongly and independently 
    of the radial 
    mode distance -- i.e.\ gradient driven double-resonance works even if there is no radial mode overlap.  
    In contrast, the enhancement of the inner and weaker RSAE decreases with the radial 
    mode distance for small $\Delta s$. Then it increases again for $\Delta s > 0.15$. 
    These larger mode distances match the double-resonant
    particle orbits and therefore enable inter-mode energy exchange, driving the weak mode. For
    higher $\Delta s$, the larger, i.e.\ higher energetic orbits fit the mode distance
    and lead to even more energy exchange.
    Furthermore, with larger radial mode distances, the modes are able to tap energy from a wider 
    gradient region. The amplitude ratios, however, are even lowered in the double-resonant case 
    compared to the respective single mode levels, if the radial mode distance is small. This happens 
    due to the mutual gradient depletion at the 
    other mode's radial position. If the modes feature a larger radial distance ($\Delta s > 0.15$), the
    double mode scenario amplitudes are much higher compared to the single mode amplitudes, both for the
    TAE and the RSAE. Both modes benefit from each other -- most for a radial distance of about 
    $\Delta s \approx 0.25$.
    It is important to note that the distance $\Delta s=0.25$ giving maximum amplitude ratios depends
    strongly on the absolute mode positions with respect to the radial distribution function, 
    and especially on the amplitude regime (stochastic or non-stochastic) of each mode. 
    The same applies for the value $\Delta s=0.15$ at which the transition towards
    double-resonant amplitude enhancement takes place.\\
    To summarize, one can say, that growth rates are
    generally enhanced by the presence of another mode, whereas for the amplitudes, this is only
    the case, if the modes feature a sufficient radial distance. For small distances, modes at radial 
    positions, where the initial distribution function is already relatively flat, double-resonance 
    leads to strong mode stabilization. If the amplitudes are enhanced, their amplification level is, 
    however, mainly determined by whether the mode reaches the stochastic regime. 
    The probability of reaching the stochastic threshold rises strongly if a second mode is present.
    However, if the growth rates are relatively low for some reason 
    (e.g.\ small mode width, small fast particle beta), the particle redistribution 
    is not strong enough to lead to a dominant gradient driven double-resonance. 
    In these cases, the \textit{inter-mode energy transfer} mechanism can prevail
    (even in a later phase). The dominant mode is weakened through inter-mode energy transfer
    to the subdominant mode,
    as depicted in \fref{run0186+0187+0188_amplitude}. The process saturates, as
    the modes' amplitudes converge towards comparable levels.
    In this simulation, there is no strong fast particle redistribution.
    In the monotonic \qp\ equilibrium (shown in \fref{q23824}),
    the Alfv{\'e}nic modes as simulated within this study do not grow
    up to the stochastic regime. In order to also be able to investigate
    the stochastic phase in the monotonic profile case, one can shift the modes further outward radially,
    where the mode drive is higher. Such simulations revealed interesting possibilities for interplay
    between the two double-resonance mechanisms: e.g., it is possible that the mode in the 
    flattened-gradient region is not weakened, but rather the \emph{other} mode. 
    In this scenario the mode at the position
    of the flattened gradient is driven solely through energy transfer
    from the other mode, as there is no gradient-caused mode drive remaining.
    Eventually, this leads to a depletion of the other (stronger) mode's amplitude.

    \subsubsection{The role of trapped particles}
    Another question regards the role of trapped particles in the double-resonance mechanisms:
    Especially for the inter-mode energy exchange, the particles' orbit 
    widths are crucial. Therefore, the
    question arises if the major part of the double-resonance is carried 
    by the trapped particle population or by the passing particles.\\
    In \textsc{Hagis} it is possible not to load any trapped particles at all. In 
    the simulations presented in the following,
    the code was modified to even neglect particles that become trapped
    in the course of the simulation.\\
    It was found that for the linear phase, the absence of trapped particles 
    leads to significantly lower growth rates. This effect is slightly stronger 
    for inner mode positions, although there would be more trapped particles further
    outside. This is due to the fact that the orbit width broadens with radial position, 
    and therefore also the loss region. But especially the resonance with a broader TAE 
    is effective only with broader orbit particles that are more easily lost.\\
    However, except for the effect of generally lower amplitudes in the scenario without trapped particles, 
    the double-resonance also works with passing particles alone: 
    the effect of double-resonance
    remains, which is reflected by the ratio of growth rates and amplitude levels in double mode simulations 
    over those from single mode simulations. This means, passing as well as banana orbits 
    can be adequate orbits for the double-resonance mechanism of inter-mode energy transfer:
    As for passing particles, the orbit does not close entirely on the low field side,
    the double-resonance mechanism must work with the higher poloidal harmonics $m$ of one mode, 
    located at the high field side. For the TAE, due to strong ballooning effects, there are no
    significant maxima at the high field side. Core-localized modes, such as 
    BAE and RSAE do not couple to the next harmonic $m+1$ and are cylindrical modes that 
    feature higher harmonics at the high field side. There, double-resonant passing particles 
    can transit the mode. Compared to the banana width $w$ of a trapped particle, the 
    drift displacement $\Delta r$ of a passing particle's orbit (at the same energy) is 
    significantly smaller. But, as the resonance condition is different for passing particles, 
    higher energetic particles are responsible for double-resonance. These feature a larger 
    drift displacement, that fits again the simulated radial mode distances $\Delta s$.\\
    However, for the gradient driven (double as well as single) resonance, the trapped particles' 
    resonances are very important. It was not possible to quantify this further.
    Without trapped particles, the saturation levels are much lower and do not reach the stochastic regime.
    This is due to the importance of the trapped particles' (single) resonances, leading to gradient driven
    mode-particle interaction. The importance of the trapped particles is consistent
    with the fact, that the TAE is an even TAE, which is known to interact very strongly with trapped particles.

    \subsubsection{The importance of equal toroidal mode numbers}
\begin{figure}[H]
      \centering
      \includegraphics[width=0.6\textwidth,height=5.5cm]{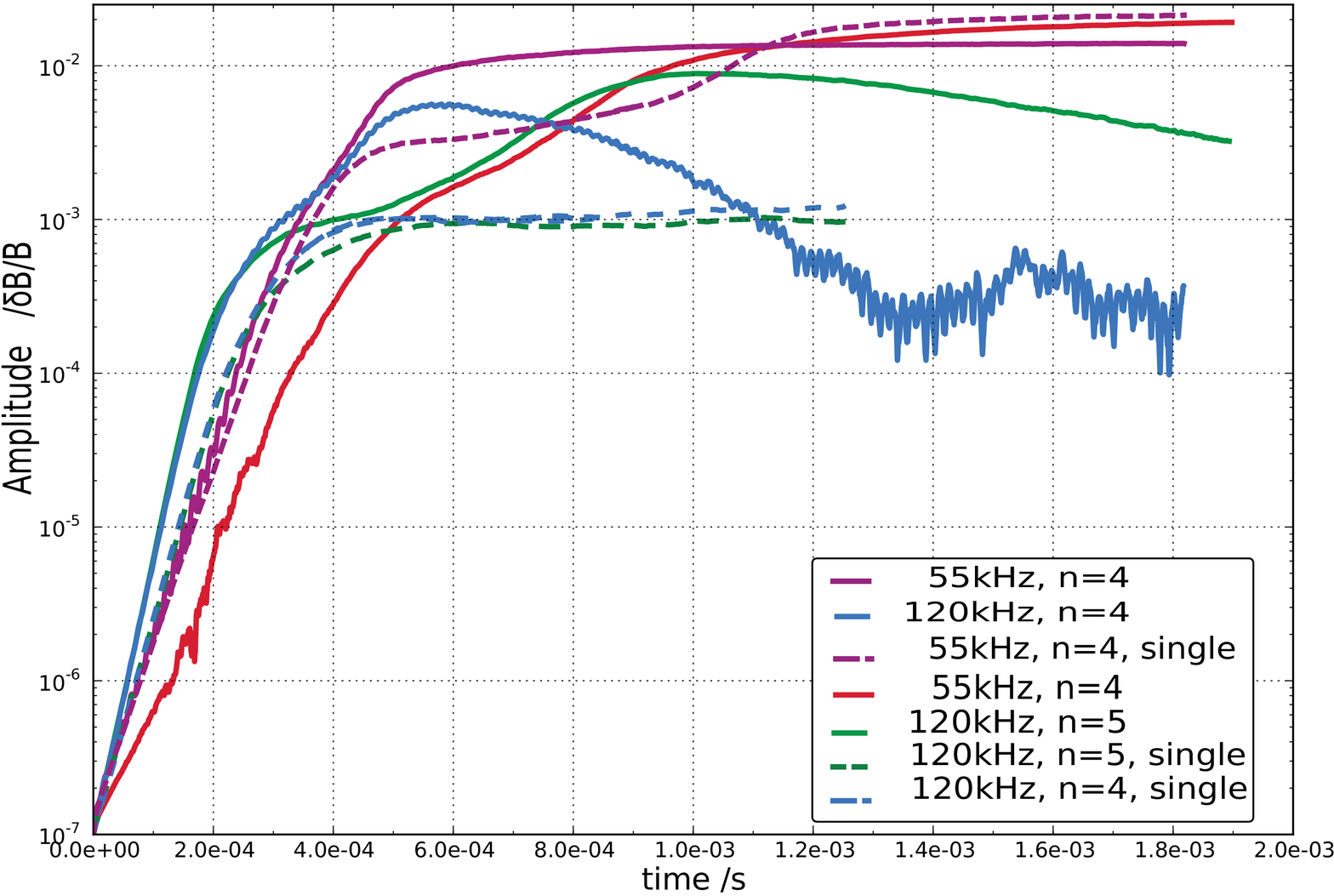}
      \caption{\itshape Mode amplitudes evolution in double mode simulations with equal
        toroidal mode numbers $n = 4$ (pink solid line: RSAE, blue solid line: TAE)
        and with different $n$: RSAE (red solid, $n=4$) and TAE (green solid $n=5$).
        For comparison, the single mode simulations are given (dashed). The RSAE is
        located at $s=0.3$, the TAE at $s=0.4$.} 
          \label{run0162+0165+0167+0227+0228_amplitude}
    \end{figure}

    Especially for the inter-mode energy transfer, there has to be a population of particles that 
    is resonant with both modes at once. For two modes with different toroidal mode numbers $n$,
    the overlapping volume of resonances in phase space becomes smaller, and it becomes 
    less probable to find particles that fulfill the resonance condition eq.\ (\ref{doublerescond}) for 
    two modes at once, and it is expected that less double mode resonance effects occur, 
    if the modes differ in their toroidal mode number $n$.
    \Fref{run0162+0165+0167+0227+0228_amplitude} shows the resulting amplitude 
    evolution in a double mode simulation
    with different $n$ for the RSAE ($n=4$) and TAE ($n=5$).
    In the linear phase, one can see clearly that the amplitudes of the single TAE 
    simulations are almost exactly equal, independent on $n$. However, in the double mode scenario,
    the RSAE is enhanced (compared to the single mode amplitude), and the TAE 
    even more strongly -- but only
    if both modes have equal $n$. If this is not the case, only the TAE is enhanced, whereas the 
    RSAE actually decreases. Furthermore, the superimposed oscillation 
    vanishes to a slightly jagged curve. This shows that gradient driven
    double-resonance still works with different toroidal $n$ -- the outer mode amplitude is still
    enhanced -- but the inter-mode energy transfer mechanism breaks down, as the resonance 
    condition cannot be easily fulfilled for both modes at once for particles with trajectories
    through both modes: the inner RSAE amplitude is not enhanced, but weakened.
    When reaching saturation, the effect of inter-mode energy transfer is overlapped by
    the dominance of the gradient driven double-resonance, and therefore, the saturation levels do 
    not depend strongly on equal $n$. Simulations with different $n$ show almost identical nonlinear behavior,
    although saturation is reached later. The stochastic threshold does not seem to be influenced by 
    the differing $n$, nor the effect that one mode can be damped by the 
    redistribution of the other one. However, the differences again depend on the exact
    scenario.

\section{Conclusions}
    The interaction of fast particles with Alfv{\'e}n Eigenmodes of different frequencies was 
    studied numerically with the \textsc{Hagis} Code  
    in a simple, but physically realistic picture for two different MHD equilibria occurring 
    during the ASDEX Upgrade discharge \#23824.
    Double-resonant mode drive was compared to single mode scenarios, verifying previous 
    findings of double-resonance mechanisms:
    gradient driven double-resonance \cite{Berk92} and inter-mode energy transfer \cite{mwb_phd}.
    The latter was observed to prevail only in low amplitude cases, 
    enhancing the weaker mode at the expense of the dominant one.
    A superimposed beat frequency oscillation on both modes' amplitudes was found, 
    as well as higher linear growth rates -- at least for one mode -- compared to the single mode 
    reference cases. 
    The growth rate enhancement was observed to have only a very weak dependence  
    on the radial mode distance.
    Concerning the amplitudes of the Alfv{\'e}n Eigenmodes, 
    double-resonance can enhance modes into the stochastic regime and therefore lead to much higher 
    saturation levels compared to the single mode scenarios, even if there
    is no radial mode overlap.
    However, close radial mode distances were found to destroy
    linearly dominant modes when the nonlinear regime sets in. 
    As a consequence, linearly weaker modes may become nonlinearly dominant.
    These results reveal a complex nonlinear evolution of multi-mode scenarios, rather
    than merely a simple continuation of the linear multi-mode behavior.\\\\
    The authors wish to thank D.R.\ Hatch for proof reading.
\bibliographystyle{iopart-num}
\bibliography{literature}

\end{document}